\documentclass[twocolumn,prb,superscriptaddress,showpacs,showkeys]{revtex4-1}
\usepackage{graphicx}
\usepackage{dcolumn}
\usepackage{bm}
\usepackage{textcomp}
\usepackage{sidecap}

\newcommand{\invang}{\,\r{A}$^{-1}$}
\newcommand{\microns}{\,$\mathrm{\mu}$m}
\def\mathbi#1{\textbf{\em #1}}
\begin{document}


\title{Neutron scattering study of the high energy graphitic phonons in superconducting  CaC$_6$} 

\author{M. P. M. Dean}
\email{mdean@bnl.gov}
 \affiliation{Cavendish Laboratory, University of Cambridge, JJ Thomson Avenue, Cambridge CB3 0HE, United Kingdom}
 \affiliation{Department of Condensed Matter Physics and Materials Science,
Brookhaven National Laboratory, Upton, New York 11973, USA}
\author{A. C. Walters}
    \affiliation{European Synchrotron Radiational Facility, Polygone Scientifique Louis N\'{e}el, 6 rue Jules Horowitz, 38000 Grenoble, France}

\author{C. A. Howard}
    \affiliation{London Centre for Nanotechnology and Department of Physics and Astronomy, University College London, London WC1E 6BT, United Kingdom}

\author{T. E. Weller}
    \affiliation{Harvard University Biological Laboratories, 16 Divinity Avenue, Cambridge, MA 02138-2020, USA}

\author{M. Calandra}
    \affiliation{Universit\'{e} Pierre et Marie Curie, 4 Place Jussieu-case postale 115, 72252 Paris cedex 05, France}

\author{F. Mauri}
    \affiliation{Universit\'{e} Pierre et Marie Curie, 4 Place Jussieu-case postale 115, 72252 Paris cedex 05, France}

\author{M. Ellerby}
\affiliation{London Centre for Nanotechnology and Department of Physics and Astronomy, University College London, London WC1E 6BT, United Kingdom}

\author{S. S. Saxena}%
 \affiliation{Cavendish Laboratory, University of Cambridge, JJ Thomson Avenue, Cambridge CB3 0HE, United Kingdom}

\author{A. Ivanov}
    \affiliation{Institut Laue-Langevin BP 156, Polygone Scientifique Louis N\'{e}el, 6 rue Jules Horowitz, 38000 Grenoble, France}

\author{D. F. McMorrow}
\affiliation{London Centre for Nanotechnology and Department of Physics and Astronomy, University College London, London WC1E 6BT, United Kingdom}

\date{\today}

\begin{abstract}
We present the results of a neutron scattering study of the high energy phonons in the superconducting graphite intercalation compound CaC$_6$. The study was designed to address hitherto unexplored aspects of the lattice dynamics in CaC$_6$, and in particular any renormalization of the out-of-plane and in-plane graphitic phonon modes. We present a detailed comparison between the data and the results of density functional theory (DFT). A description is given of the analysis methods developed to account for the highly-textured  nature of the samples. The DFT calculations are shown to provide a good description of the general features of the experimental data. This is significant in light of a number of striking disagreements in the literature between other experiments and DFT on CaC$_6$. The results presented here demonstrate that the disagreements are not due to any large inaccuracies in the calculated phonon frequencies.

\end{abstract}


\maketitle

\section{\label{sec1}Introduction}
Superconductivity has been known in graphite intercalation compounds (GICs) for over 40 years.\cite{Hannay1965} Until recently, however, the maximum ambient pressure superconducting transition temperature was limited to about 2\,K.\cite{EnokiBook} The discovery of superconductivity in CaC$_6$ at the enhanced transition temperature $T_c$ = 11.5\,K changed this and reignited the interest in superconductivity in graphite intercalation compounds (GICs).\cite{Weller2005,Emery2005a} Although early reports suggested that the superconducting mechanism involved coupling via acoustic plasmons,\cite{Csanyi2005} the current consensus is that the superconductivity is close to \emph{s}-wave symmetry\cite{Lamura2006,Sutherland2007,Sanna2007} and phonon mediated.\cite{Calandra2005,Mazin2005,Kim2007} Having said this, some controversy remains as different experimental and theoretical techniques implicate different phonons and electrons as being primarily responsible for the coupling. Experiments have revealed a Ca isotope coefficient of 0.5, which is consistent with coupling to only the low energy Ca phonons.\cite{Hinks2007} Density functional theory (DFT), however, predicts that in-plane Ca$_{xy}$ vibrations and out-of-plane C$_z$ phonons contribute roughly equally, coupling electrons via transitions involving the interlayer band.\cite{Calandra2005} Information about electron-phonon coupling can also be inferred from specific heat of CaC$_6$ measurements.\cite{Kim2006specheat} A detailed comparison between the experimental results and calculations \cite{,Mazin2007} finds that the electron-phonon coupling to the C$_z$ modes is stronger than DFT predicts. Further adding to the inconsistencies, a recent angle-resolved photo-emission (ARPES) measurement proposes a entirely different picture, as it finds sufficient coupling between the in plane C$_{xy}$ modes and the $\pi^*$ electrons to account for T$_c$ without other contributions.\cite{Valla2009} However, the extraction of this coupling is made difficult as the bands are predicted by DFT to curve in that region.\cite{Calandra2007} A first step to resolving these problems is to measure the phonons to ensure that they are accurately predicted by DFT.


Due to the highly anisotropic structure of CaC$_6$, its phonons can be divided into different groups based on the predominant atomic motion, which occupy different energy windows. The Ca atoms are much heavier and more weakly bonded than the C atoms, so the Ca-related phonons occur at the lowest energies typically below $\sim$ 40\,meV. The graphene sheets have strong planar bonding, meaning that the out-of-plane C$_z$ modes fall between 40-80\,meV, while the in-plane C$_{xy}$ modes are higher energy and dominate the spectrum from 100-180\,meV. To date the main phonon dispersion measurements in CaC$_6$ were performed using inelastic x-ray and neutron scattering and concentrate on phonons below $\sim$ 40\,meV.\cite{Upton2007,Astuto2009}. Although some phonon modes of the measured dispersion are well reproduced by the DFT calculations of Refs. \cite{Calandra2005,Calandra2006a}, other modes are substantially underestimated in energy.\footnote{A. C. Walters Private Communication} Furthermore, an anti-crossing of a longitudinal acoustic mode with an unknown mode was discovered,\cite{Astuto2009} which was tentatively ascribed to vacancies in the CaC$_6$ crystal. The higher energy phonons in CaC$_6$ have only been measured using Raman scattering.\cite{Dean2010,Hlinka2007,Mialitsin2009} Raman scattering normally only measures phonons at small momentum transfer near the center of the Brillouin zone. Furthermore, phonons measured by Raman scattering are not representative of the phonons across the Brillouin zone, as they are subject to non-adiabatic effects, which lead to large energy shifts at the zone center.\cite{Dean2010,Saitta2008}

For these reasons we have performed a neutron scattering measurement of the high energy C$_{xy}$ and C$_z$ phonons in CaC$_6$. There are two principle aims: to compare experiment measurement and theoretical predictions for the high energy phonons in CaC$_6$ and to search for any temperature dependence of the phonons, which if present, might imply strong electron-phonon coupling. We note that this is the first neutron scattering measurement of the high energy graphitic phonons in any GIC. Section \ref{sec:expmethods} describes the experimental methods. Macroscopic GIC samples are typically highly textured, due to the texture of the starting graphite, which means that an average of the scattered intensity is performed  over a large swath of reciprocal space when the neutron spectrometer is set for nominal energy and wavevector transfers. Section \ref{sec:thmethods} outlines calculations to account for the sampling of the Brillouin zone to allow a detailed comparison between DFT and experiment. We discuss the good overall agreement between experiment and theory and how it could be further improved in section \ref{sec:results}. Section \ref{sec:conclusions} outlines the implications of our work in light of the current literature on CaC$_6$.

\section{Experimental Methods\label{sec:expmethods}}
Samples were synthesized using the Li-Ca alloy method from $\sim$ 50 strips of highly orientated pyrolytic graphite (HOPG).\cite{Pruvost2004} In this method the Li atoms initially diffuse into the graphite to form LiC$_6$, opening the graphite layers, after which the Ca is able to replace the Li to form the final sample. The resulting 700\,mg of sample was sealed within an aluminium can to protect it from reacting with air, with the strips mounted on an aluminium plate with the sample c-axes aligned with one another. Thin aluminium foil was used to hold the sample in place, to avoid the possibility of glue reacting with the sample. The sample was measured using IN1 at the Institute Laue-Langevin (ILL). This instrument is connected to the hot source of the ILL reactor, which provides neutrons of sufficiently high energy to access the 200\,meV energy transfers required in this study. A Cu (2 2 0) monochromator was chosen and scans were performed at a fixed final energy of 100\,meV, and an erbium filter was inserted before the double-focusing Cu (2 0 0) analyzer to absorb spurious elastic signals.\cite{ILLwebsite} The sample was loaded into a closed-cycle refrigerator to control the sample temperature. Data were taken at 5 and 20\,K and summed, taking into account changes in the Bose-Einstein factor, as these spectra proved to be identical. In this experiment we use the hexagonal representation of the CaC$_6$ unit cell with $a$ = $b$ = 4.333\,\r{A}, $c$ = 13.572\,\r{A}, $\alpha$ = $\beta$ = 90$^{\circ}$ and $\gamma$ = 120$^{\circ}$, because this distinguishes the quasi-two dimensional nature of the sample more clearly than the rhombohedral unit cell. Due to the nature of the starting graphite (HOPG), the sample consists of crystallites of size $\sim$ 1\microns,\cite{Dresselhaus2002} which are aligned in the c-direction and powder-like in the ab-plane. Only two components are therefore required to describe the scattering vector $\mathbi{Q}$: $Q_r$ and $Q_z$ which are the in-plane and out-of-plane components measured in reciprocal lattice units.
\begin{figure}
\includegraphics[width=0.5\textwidth]{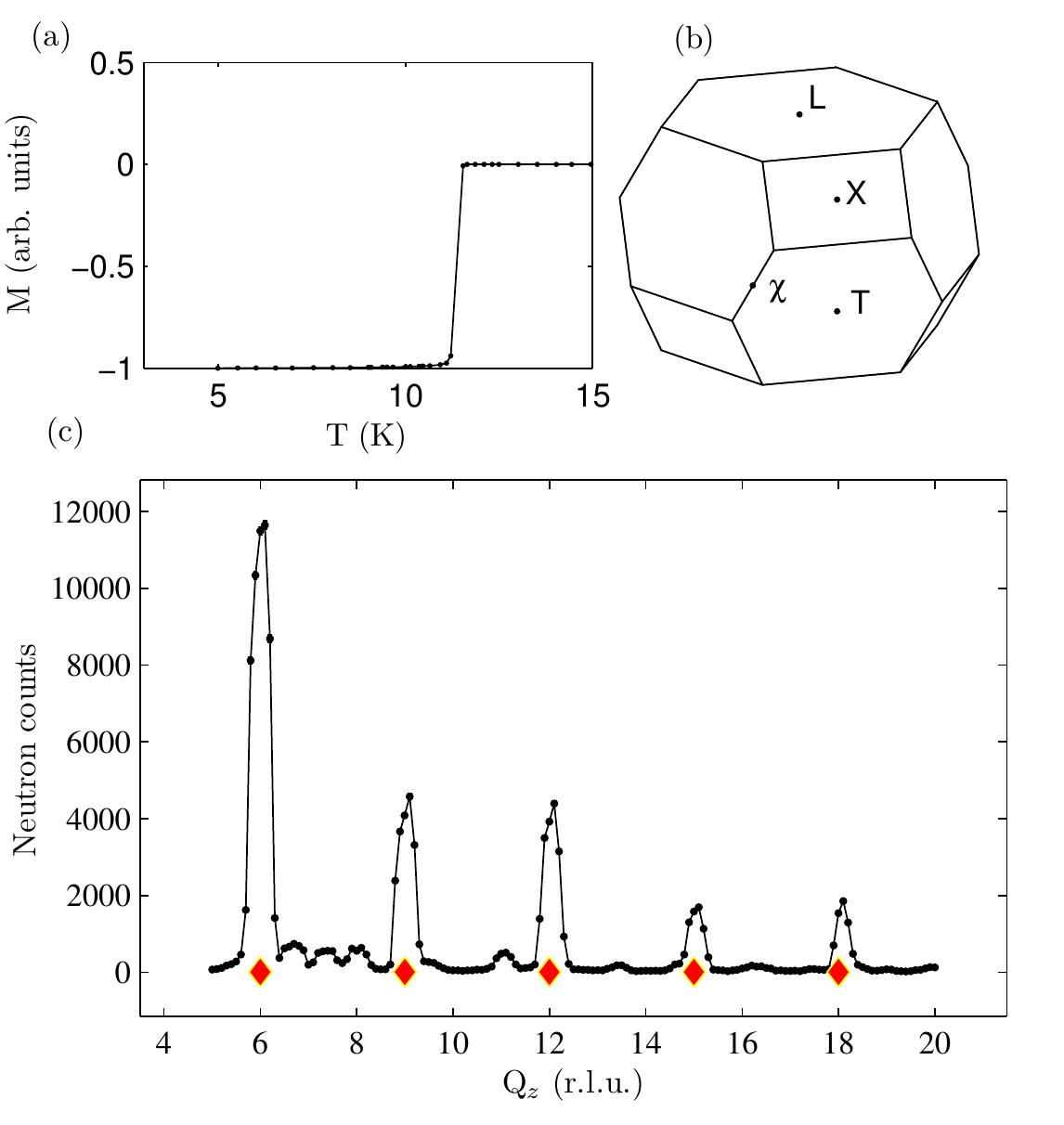}
\caption{\label{fig:diffraction} (Color online) (a) A measurement of magnetization $M$ against temperature $T$ displaying the sharp superconducting transition in CaC$_6$. (b) The Brillouin zone of CaC$_6$. (c) A (0 0 L) diffraction scan on the sample measured using IN1. Tick marks denote the positions of the CaC$_6$ Bragg peaks.}
\end{figure}

Figure \ref{fig:diffraction}(c) shows a $Q_z$ (0 0 L) elastic diffraction scan performed on the sample using IN1. All the principal peaks are indexed to CaC$_6$ denoted by the markers. Only very small peaks are seen coming from the aluminium can and LiC$_6$ and graphite impurities. The graphite and LiC$_6$ impurities are estimated to each account for $\approx$ 5\,\% of the sample by volume, which is consistent with a number of published works. \cite{Hinks2007,Kim2007} This is further backed up by Fig. \ref{fig:diffraction} (a), which displays a sharp superconducting transition at 11.5\,K obtained from a sample made by the same method. A rocking curve performed on the (0 0 15) Bragg peak shows that the c-axis mosaic of the total sample can be described by a Gaussian ($e^{-x^2/2\sigma^2}$) with $\sigma$ = 4.7\,$^{\circ}$. Due to the inelastic cross section being inversely proportional to the large energy transfer,\cite{ShiraneBook} the neutron flux at the detector is relatively small and long counting times (1-2 days)  were required to obtain sufficient statistics for the analysis.

It is instructive to consider the relative neutron scattering efficiency of the C and Ca atoms within the sample, which have scattering cross sections $\sigma_{C}$ = 5.55 and $\sigma_{Ca}$ = 2.78\,barns and masses $M_{C}$ = 12 and $M_{Ca}$ = 40 atomic mass units. The phonon scattering intensity is directly proportional to the cross section and inversely proportional to the mass. Consequently when the sample stoichiometry is accounted for, the ratio of C to Ca phonon scattering is  $6 \sigma_{C} M_{Ca}/ ( \sigma_{Ca} M_{C})$ $\approx$ 40. Such a high ratio implies that the vast majority of the scattering will be from the phonons involving predominantly C atom motion.

The selection rule for neutron scattering from phonons can be described by a dot product $(\mathbi{Q}\cdot \mathbi{e})^2$, where $\mathbi{e}$ is the polarization of the phonon.\cite{ShiraneBook}  Two reciprocal space locations were chosen to take data at with high statistics: $Q_r$ = 6, $Q_z$ = 0 was employed to study the C$_{xy}$ phonons between 80 and 200\,meV and $Q_r$ = 2 $Q_z$ = 24 was chosen to study the C$_z$ phonons between 20 and 100\,meV. A preliminary measurement at $Q_r$ = 0 $Q_z$ = 24 showed relatively little phonon structure, so this $\mathbi{Q}$ was not chosen for the final measurement.

\begin{figure}[t]
\includegraphics[width=0.5\textwidth]{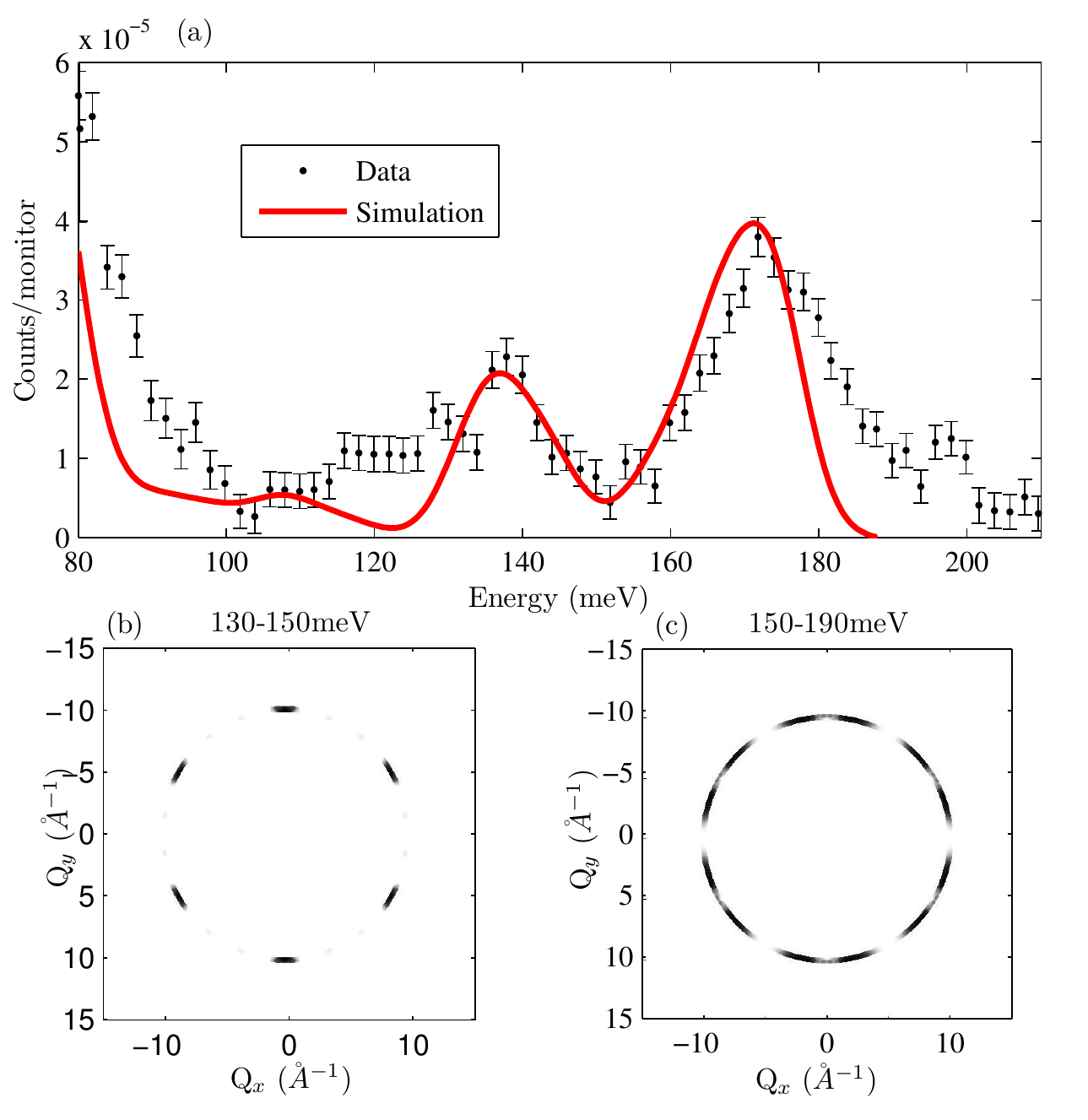}
\caption{\label{fig:600} (Color online) Comparison of the neutron scattering measurements and theoretical simulations of the C$_{xy}$ phonons in CaC$_6$ measured at $Q_r$ = 6, $Q_z$ = 0 from 80-200\,meV. (a) The neutron scattering signal. Black points represent the measured spectrum and the solid line is the result of the simulation. An exponential background has been subtracted from the data. (b) A cut through reciprocal space in the $Q_z$ = 0 plane, which depicts the regions that contribute most strongly to the peak in the phonon intensity simulation in panel (a) from 130-150\,meV. Black represents the strongest contribution and white represents no contribution. (c) A similar calculation to (b) for the second peak in the data in the energy region 150-190\,meV. The spectra are the summed result of scans performed at 5 and 20\,K.}
\end{figure}

\section{Theoretical Methods\label{sec:thmethods}}
The textured nature of the sample presents a challenge for comparison between theory and experiment. Detailed calculations were therefore required to facilitate this comparison. The first step is to compute neutron spectrum for arbitrary $\mathbi{Q}$. Phonon dispersion were calculated using DFT in the linear response, as implemented in the QUANTUM-ESPRESSO code\cite{QE-2009,Baroni2001}, using the generalized gradient approximation \cite{Perdew1996} and ultrasoft pseudopotentials.\cite{Vanderbilt1990} The dynamical matrix was calculated on a $4\times 4\times 4$ k-point grid and then Fourier interpolated at any point in reciprocal space. Full details are given in Refs.  \onlinecite{Calandra2005,Calandra2006a}. The Fourier interpolated dynamical matrix at a given $\mathbi{Q}$ vector is then diagonalized and the neutron structure factor is calculated along the lines of Ref. \onlinecite{Astuto2007} using neutron form factors and a Gaussian resolution, compatible with neutron scattering experiments.
To account for the texture of the sample the structure factor is calculated at several points included in a toroidal-like volume in reciprocal space, which is most straightforwardly described in spherical polar coordinates $(|\mathbi{Q}|,\theta,\phi)$.
As the sample is powder-like in plane this causes all $\theta$ to be sampled equally. The c-axis mosaic, as determined by the rocking curve, is the dominant contribution to the sampling of $\phi$, which is represented using Gaussian sampling with $\sigma$ = 4.7\,$^{\circ}$. The sampling of $|\mathbi{Q}|$ is described by another Gaussian with $\sigma$ =  0.1\invang, which is consistent with the width of the in-plane diffraction peaks.  This volume was sampled randomly and about 1000 $\mathbi{Q}$-points were required to achieve good convergence. The instrumental resolution function was calculated using the RESTRAX software package,\cite{Saroun1997} which performs Monte Carlo neutron simulations of the combined effect of IN1 and the sample. This function was convolved with the simulated spectra. Each simulation was checked to confirm that it was stable to small changes in the mosaic parameters.

\begin{SCfigure*}
\includegraphics[width=0.6\textwidth]{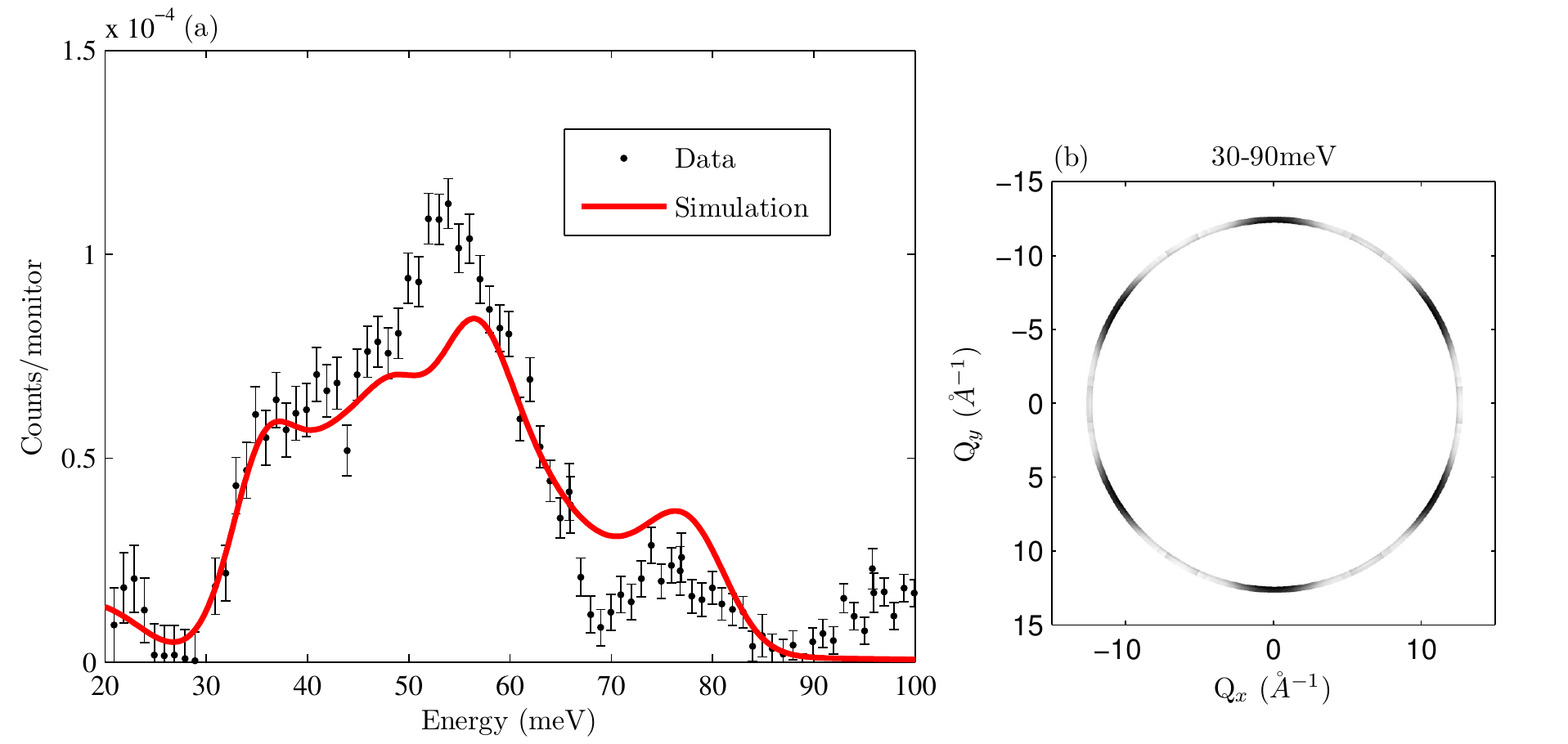}
\caption{\label{fig:2024} (Color online) (a) The inelastic neutron scattering signal for the C$_z$ modes in CaC$_6$ measured between 20-100\,meV at $Q_r$ = 2 $Q_z$ = 24. Black points represent the measured spectrum and the solid line is the result of the simulation. An exponential background has been removed from the data to account for the incoherent scattering from the Li impurities and the aluminium can. (b) A simulated section of reciprocal space in the $Q_z$ = 24 plane, which depicts the regions of reciprocal space which contribute most strongly to the peak in the phonon intensity simulation in panel (a). The spectra are the summed result of scans performed at 5 and 20\,K.}
\end{SCfigure*}

\section{Results\label{sec:results}}
The C$_{xy}$ modes were measured with $Q_r$ = 6, $Q_z$ = 0 and are depicted by the black points in Fig. \ref{fig:600}(a), which can be compared to the solid line from the theoretical simulation. 
We first discuss the two main peaks which dominate the spectrum at 137 and 172\,meV respectively. These are reasonably well reproduced by our first principles calculations. In particular, the peak at 137\,meV is reproduced well in the simulation and the peak at 172\,meV is predicted to be narrower and at about 2\,meV lower energy. Further calculations were performed to analyze the phonons contributing to these peaks. Figure \ref{fig:600}(b) and \ref{fig:600}(c) indicate the relative contributions to the scattering intensity for the two principle phonon peaks in the ranges 130-150 and 150-190\,meV respectively. The calculations reveal that different regions of the Brillouin zone provide the majority of the phononic signal.  The strong variation in scattering efficiency from various small areas of the Brillouin zone demonstrates why a relatively large number of sampling points ($\sim$ 1000) were required to adequately represent the Brillouin zone. From Fig. \ref{fig:600}(b) one can see that the $\sim$\,137\,meV peak comes predominantly from $\mathbi{Q}$ = (6 0 0) direction and symmetry related points.  We note that the six-fold symmetry of these phonons arises because for the C based modes, the influence of the intercalant superlattice is only a weak perturbation of the six-fold symmetry of the graphene layers. In this experiment a large fraction of the Brillouin zone is sampled in a rather uneven way, making the measurement almost, but not quite, a 2D density of states measurement. The intensity contributing to the peak at 150-190\,meV from is shown in Fig.\ \ref{fig:600}(c) this intensity  comes from a larger region of reciprocal space and also displays six-fold symmetry.

There are also two minor peaks at 120 and 196\,meV. The peak at 196\,meV is well above the highest calculated phonon frequency in CaC$_6$.\cite{Calandra2005,Calandra2007, Kim2007} Since our samples have roughly 5$\%$ of graphite, this peak could be due to the in-plane phonon vibrations of pristine graphite sheets, which are indeed near this energy.\cite{Mohr2007} The broad peak at 120\,meV is also not completely reproduced in theoretical calculations. A peak is present in the theoretical structure factor but at substantially lower energy (113\,meV) and with a lower intensity. It is important to note that the lower intensity is due to the highly dispersive nature of the C$_{xy}$ phonon dispersion
in this region. As stated before, the in-plane averaged structure factor is related to the 2D phonon density of states. In graphite the C$_{xy}$ phonons are very flat near the $K$ point at $\approx 120$ meV. Consequently this peak could also be due to pristine graphite. It would be surprising, however, if a $5\%$ content of graphite could give a relatively significant signal in neutron scattering.

The C$_z$ modes within CaC$_6$ were measured at $Q_r$ = 2 and $Q_z$ = 24, where the large $Q_z$ component acts to enhance the scattering power of the \emph{z}-polarized phonons. In this region of the spectrum, the strong incoherent scattering from the Li impurities and the signal from the aluminium caused a background, which was modeled as two exponential functions and subtracted from the data. The data has peaks at 54 and 76\,meV. It can be seen from Fig. \ref{fig:2024}(a) that the simulated spectrum reproduces the general shape of the spectrum quite well, however there are some differences. Although the position of the peak at 76meV is predicted, its relative intensity is underestimated. Furthermore both the position in energy and the relative intensity of the main peak at 54\,meV are overestimated. The fact that the 54\,meV peak is overestimated in energy at 56.4\,meV is especially interesting as the energy of the C$_z$ modes are sensitive to the occupation of the $\pi^*$ band: as the $\pi^*$  band is populated these mode soften.\cite{Boeri2007} Furthermore, a softening of the Raman active C$_z$ phonon at $\mathbi{q}$ $\approx$ 0 has been shown, to correlate with the increase in $T_c$ in BaC$_6$, SrC$_6$, YbC$_6$, to CaC$_6$.\cite{Dean2010} This mode softens from 60\,meV in BaC$_6$ to 54\,meV in CaC$_6$. The measurement of the energy of this optical phonon in this study is thus important and could suggest that the amount of charge in the $\pi^*$ band is underestimated by the DFT model. Further inelastic neutron scattering or inelastic x-ray scattering experiments measuring the C$_z$ phonons could be particularly interesting. Fig. \ref{fig:2024}(b) displays the analysis performed for this scan. The majority of the neutron scattering intensity from 30-90\,meV comes from the (0 2 24) direction and the five other symmetry related points.

The temperature dependence of a phonon spectrum can often provide information regarding the origin of the phonon linewidth. For example, electron-phonon coupling such as that seen in Ref.\ \onlinecite{Valla2009} typically leads to temperature-independent linewidths, while anharmonic effects usually lead to a reduction in linewidth at low temperatures. Studying such a change in linewidth using the triple axis mode of IN1 is rather limited by the flux of the instrument. So the scans in Figs. \ref{fig:600} and \ref{fig:2024} could only be performed at 5 and 20\,K, which proved identical. As an alternative approach we chose to look for temperature dependent effects in the phonon spectrum by operating IN1 in the Be filter mode. In this mode a block of beryllium is placed between the sample and the detector, this absorbed all neutrons with  energy $>$ 5.2\,meV. The large aperture means that a large proportion of $\mathbi{Q}$ is sampled and a larger neutron flux is registered at the detector, making a complete temperature dependent study practical.

\begin{figure}[b]
\includegraphics[width=0.50\textwidth]{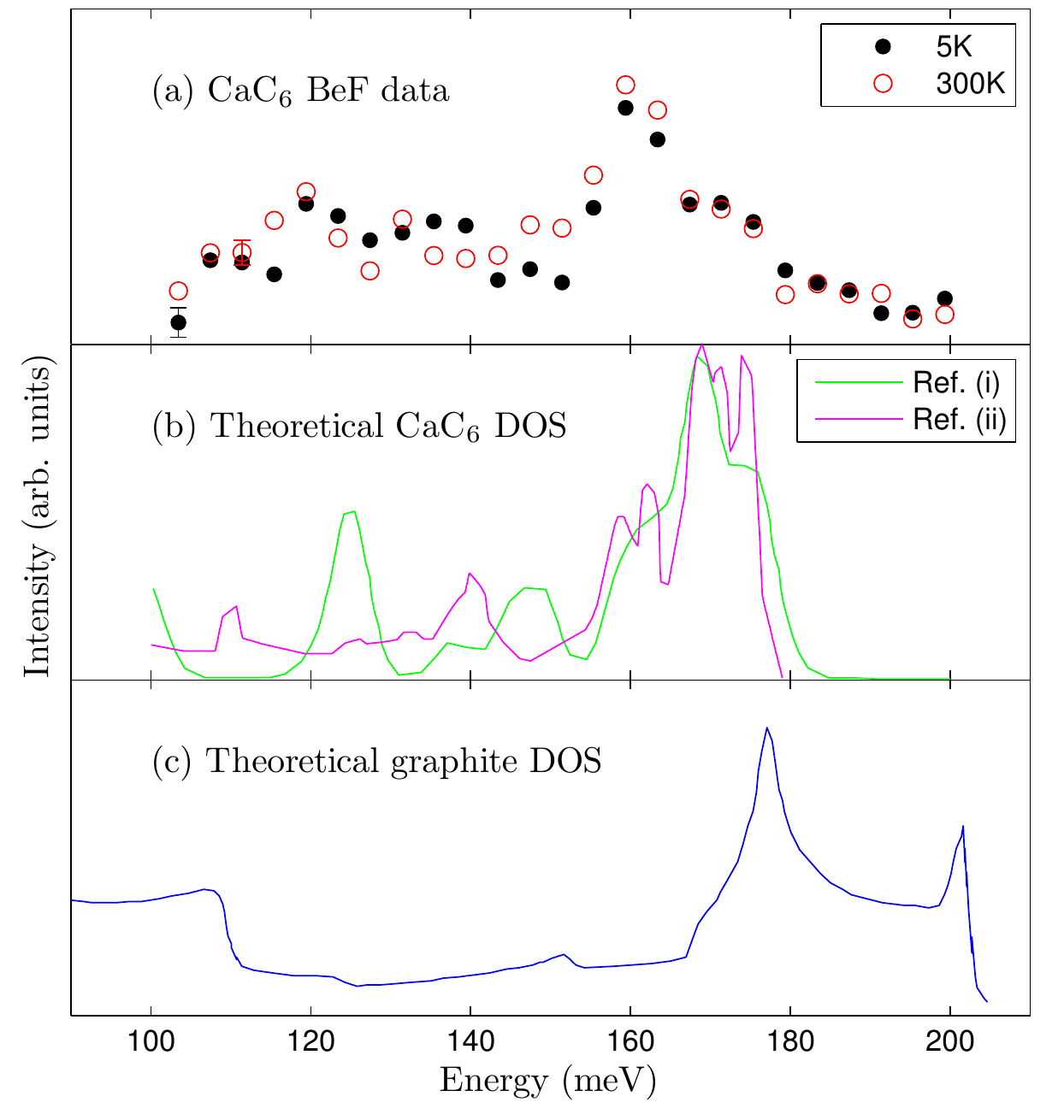}
\caption{\label{fig:PhDOS} (Color online) (a) The beryllium filter measurement of the C$_{xy}$ phonons in CaC$_6$ at 5\,K ($\bullet$) and 300\,K ($\circ$), the error bars shown are representative of the error bars on all points. (b) The theoretical CaC$_6$ phonon density of states (DOS) taken from (i) Calandra and Mauri\cite{Calandra2005} and (ii) Kim et al.\cite{Kim2006pressure}. (c) The theoretical phonon DOS of graphite \cite{Wirtz2004} shown for comparison with CaC$_6$.}
\end{figure}

 The beryllium filter spectrometer was used to measure the C$_{xy}$ phonons at a range of temperatures between 5-300\,K. Figure \ref{fig:PhDOS}(a) displays two measurements at the extremes of the temperature range 5 and 300\,K, where the spectra has been corrected for the small changes due to the Bose-Einstein factor in the neutron scattering cross section. The error bars shown are representative of the error bars on all points. Within the experimental error there is no significant change to the spectrum. A much higher resolution study using, for example, inelastic x-ray scattering might prove a fruitful method to search for any more subtle changes in the spectrum. For this measurement, the sample was oriented such that $\mathbi{Q}$ was approximately parallel to the c-axis, however, due to the large aperture significant in-plane $\mathbi{Q}$ was also sampled. The measured spectrum has strong similarities to the published density of states calculations from Refs. \onlinecite{Calandra2005,Kim2006pressure} reproduced in Fig. \ref{fig:PhDOS}(b), but the uneven $\mathbi{Q}$ sampling leads to some phonon modes appearing more strongly in the experimental measurement. This means that this measurement cannot distinguish between the two CaC$_6$ phonon calculations in the literature (Refs. \onlinecite{Calandra2005,Kim2006pressure}). In Fig. \ref{fig:PhDOS}(c) the calculated phonon density of states of graphite \cite{Wirtz2004} is plotted for comparison with CaC$_6$. We see that the main peak in graphite is at higher energy than the peak in CaC$_6$, because the Ca atoms act to dope the graphene sheets in CaC$_6$ which increases the C-C bond length\cite{Pietronero1981} and softens the phonons. Other changes to the graphene sheets, such as rolling the sheets to make carbon nanotubes, have a much weaker effect on the high energy phonons. For example, the phonon density of states for single walled carbon nanotubes the 150-200\,meV energy range is similar to that of graphite.\cite{Rols2000}

\section{Conclusions\label{sec:conclusions}}
We have performed inelastic neutron scattering experiments to measure the high energy C$_z$ and C$_{xy}$ phonons in CaC$_6$. Due to the highly textured nature of these samples, detailed calculations were required to make the comparison between experiment and density functional theory (DFT) predictions. Despite the differences discussed, overall for both the C$_{xy}$ and the C$_z$ phonons we find reasonable agreement between experiment and theory. This shows that DFT predicts the phonon structure of CaC$_6$ reasonably well, which is significant in light of several conflicting papers \cite{Calandra2005,Mazin2005,Kim2007,Hinks2007,Valla2009}. The measurements presented here imply that experimental and theoretical studies need to readdress the electronic structure and electron-phonon interaction to definitively explain superconductivity in CaC$_6$.
\begin{acknowledgments}
We thank the EPSRC, the Institut Laue Langevin, Selwyn College, Cambridge and Jesus College, Cambridge for funding. We also thank M.\ d'Astuto and G.\ Loupias for discussions. Calculations were performed at the IDRIS supercomputing center (project 081202). The work at Brookhaven is supported in part by the US DOE under Contract No. DE-AC02-98CH10886 and in part by the Center for Emergent Superconductivity, an Energy Frontier Research Center funded by the US DOE, Office of Basic Energy Sciences.
\end{acknowledgments}

\end{document}